\documentclass[superscriptaddress,twocolumn,prb,aps,groupaddress]{revtex4-1}
\usepackage{graphicx}
\usepackage{amsmath}
\usepackage{color}

\begin{document}
 	
 	\title{Tuning exchange interactions in antiferromagnetic Fe/W(001) \\ by $4d$ transition-metal overlayers}
 	
 	 	\author{Nanning Petersen}
		\altaffiliation[Current address: ]{Max Planck Institute for Polymer Research, D-55128 Mainz, Germany}
 	\affiliation{Institut f\"ur Theoretische Physik und Astrophysik,
 		Christian-Albrechts-Universit\"at zu Kiel, D-24098 Kiel, Germany}
		
 	\author{Sebastian Meyer}
	\altaffiliation[Current address: ]{Nanomat/Q-mat/CESAM, Universit{\'e} de Li{\`e}ge, B-4000 Sart Tilman, Belgium}
 	\affiliation{Institut f\"ur Theoretische Physik und Astrophysik,
 		Christian-Albrechts-Universit\"at zu Kiel, D-24098 Kiel, Germany}
		
 	\author{Stefan Heinze}
   \email[Email: ]{heinze@physik.uni-kiel.de}
 	\affiliation{Institut f\"ur Theoretische Physik und Astrophysik,
 		Christian-Albrechts-Universit\"at zu Kiel, D-24098 Kiel, Germany}
 	
 	\date{\today}
 	
 	\begin{abstract}
	We use first-principles calculations based on density functional theory to study how the magnetic properties of an Fe monolayer on a W(001) surface -- exhibiting a $c(2 \times 2)$ antiferromagnetic ground state --
	can be modified by an additional 4\textit{d} transition-metal overlayer. 
	To obtain an overview of how the $4d$-band filling influences the exchange interactions in the Fe layer 
	we have calculated
	the energy dispersion of spin spirals for 4\textit{d}/Fe/2W unsupported quadlayers, in which the W(001) 
	substrate is represented by only two atomic layers. 
Hybridization with the overlayer leads to a reduced ferromagnetic nearest-neighbor exchange interaction and the 
next-nearest neighbor exchange gains in strength.
Surprisingly, we find that the $c(2 \times 2)$ antiferromagnetic state 
is unfavorable for all systems with a $4d$ overlayer. For $4d$ overlayers from the beginning (Nb) or end (Pd) of the 
series we find a ferromagnetic ground state. As one moves to the center of the series there is a transition via
a spin spiral (Mo, Rh) to a $p (2 \times 1)$ antiferromagnetic ground state (Tc, Ru).
We have studied the Mo, Ru, and Pd overlayer on Fe/W(001) representing the surface by a sufficiently large number
of W layers to obtain bulk like properties in its center. The energy dispersion of spin spirals show qualitatively
the same results as those from the 4\textit{d}/Fe/2W quadlayers. 
The Dzyaloshinskii-Moriya interaction calculated upon including spin-orbit coupling shows significant strength
and considerable frustration effects. The calculated magnetocrystalline anisotropy energy is large as well.
All $4d$/Fe/W(001) films are potential candidates for complex non-collinear spin structures.
 	\end{abstract}
 	
 	\maketitle
 	
 	\section{Introduction}\label{Kap: Einleitung}
 	
  Iron is the prototypical ferromagnetic element with a high Curie temperature. However, its magnetic structure can be drastically 
  modified into a non-collinear 
  spin spiral state upon changing its bulk crystal structure from bcc to fcc \cite{Tsunoda1989,Myrasov1991,Uhl1992}. In ultrathin Fe films the diversity of observed magnetic structures is even 
  larger. An antiferromagnetic (AFM) checkerboard structure was suggested \cite{Wu1992} based on density functional theory (DFT) and discovered for an Fe monolayer (ML) on the 
  W(001) surface \cite{Kubetzka2005}. Based on DFT calculations it has been predicted that a Ta$_x$W$_{1-x}$ (001) surface alloy allows to tune the state from ferro- to 
  antiferromagnetic \cite{Ferriani2007}. 
  It has been demonstrated that the nearest-neighbor exchange interaction in an Fe monolayer can be tuned from ferro- to 
  antiferromagnetic by reducing the band filling of a $4d$ or $5d$ transition-metal surface \cite{Hardrat2009}. This can lead to non-collinear magnetic ground states such as the
  N\'eel state observed for an Fe monolayer on Re(0001) \cite{Morales2016}.

  In ultrathin Fe films with a small nearest-neighbor exchange interaction intriguing magnetic ground states can occur due to the interplay with other magnetic interactions.
  For example an Fe ML on Rh(111) exhibits a double-row wise AFM (or $\uparrow \uparrow \downarrow \downarrow$) state \cite{Kroenlein2018}. Even more complex and on a nanometer scale
  is the nanoskyrmion lattice which has been found in an Fe ML on Ir(111) \cite{Heinze2011}. Atomic overlayers of $4d$ transition-metals allow to tune the    
  magnetic structure into other
  states. A Pd overlayer on Fe/Ir(111) leads to a spin spiral ground state that turns into a skyrmion lattice upon applying an external magnetic 
  field \cite{Romming2013,Dupe2014}. Depending on fcc or hcp stacking of a Rh overlayer on Fe/Ir(111)
  either a spin spiral ground state or a canted $\uparrow \uparrow \downarrow \downarrow$ state \cite{Romming2018} occurs.
	Such a change in the magnetic ground state is driven by higher-order exchange interactions as shown for a 
	Pd/Fe bilayer on Re(0001) \cite{Li2020}.
	In all of these examples there is a subtle interplay of different 
  magnetic interactions. Besides Heisenberg and higher-order exchange interactions, the Dzyaloshinskii-Moriya interaction 
	(DMI) \cite{Dzyaloshinskii1957,Moriya1960} plays a key role for non-collinear magnetic structures. 

 	The c$ (2\times 2) $-AFM ground state of Fe/W(001) is stabilized by a strong antiferromagnetic nearest-neighbor exchange interaction and
  a large magnetocrystalline anisotropy favoring an out-of-plane magnetization \cite{Kubetzka2005}.
  Experimentally no evidence for a deviation from a collinear AFM state has been observed. The DM interaction \cite{Dzyaloshinskii1957,Moriya1960} 
  which results from spin-orbit coupling (SOC) and can occur at surfaces due to the broken inversion symmetry \cite{Fert1980,Crepieux1998,Bode2007} is apparently not
  strong enough in this system 
  in comparison with exchange interaction and anisotropy to induce a non-collinear magnetic ground state. The strength of the DMI in Fe/W(001) has been obtained based
  on density functional theory (DFT) calculations \cite{Belabbes2016}. It is only a little smaller than that of Fe/Ir(111) \cite{Heinze2011}, but
  considerably weaker than for Mn/W(001) \cite{Belabbes2016} in which it induces a spin spiral ground state \cite{Ferriani2008}. 
  	
 	The exchange interaction can favor collinear magnetic states, such as the FM or the AFM state or non-collinear spin structures such as spin spirals. It can also stabilize 
  two-dimensionally modulated non-collinear spin structures such as non-chiral skyrmions \cite{Leonov2015,Lin2016}. Chiral skyrmions, on the other hand, are stabilized in ultrathin films
  due to the interfacial 
  DMI \cite{Bogdanov1989,Bogdanov2001aa,Heinze2011,Romming2013,Dupe2014,Simon2014,Leonov2015}. 
  It has been proposed that the DMI can induce 
	skyrmions in two-dimensional antiferromagnets and favorable transport properties have 
	been predicted \cite{Barker2016,Zhang2016}. 
  Since DFT provides a good description of the electronic and magnetic properties of transition-metal interfaces which 
	can host skyrmions \cite{Dupe2014,Simon2014} it can guide experimental efforts to realize antiferromagnetic skyrmions. 
  
  Here we discuss the effect of $4d$ transition-metal (TM) overlayers on the magnetic interactions in Fe/W(001) using 
	DFT. We have applied the full-potential linearized augmented plane wave (FLAPW)
  method \cite{Krakauer1979,Wimmer1981,Jansen1984} 
  as implemented in the FLEUR code \cite{FLEUR}. Fe/W(001) has been chosen since a checkerboard AFM ground state has been   
	observed \cite{Kubetzka2005} and a significant 
  DMI has been predicted based on DFT calculations \cite{Belabbes2016}. 
  However, the exchange interaction and the magnetocrystalline are apparently too strong to allow for complex non-collinear 
	spin structures such as skyrmions. 
  Based on previous studies of ultrathin Fe films we anticipate that the hybridization with a $4d$ TM overlayer can weaken 
	both interactions. 
	
	We first present the total energy difference between
  the FM and the $c(2 \times 2)$ AFM state for $4d$ TM overlayers on Fe/W(001) varying the $4d$ TM from Nb to Pd. As 
	expected the energy difference between the FM and the AFM state
  is much reduced compared to Fe/W(001). Unexpectedly, the FM state is favorable for all considered overlayers. 
  
  In order to scan a larger part of the magnetic phase space we have performed spin
  spiral calculations. We start by discussing calculations for films of four layers, denoted as quadlayers, 
  consisting of the $4d$ TM layer, the Fe layer, and two layers of the W(001) substrate. These model
  systems allow to obtain the trend of magnetic interactions and ground states. Surprisingly, we find 
  that the $p(2 \times 1)$ AFM (row-wise AFM) state is favorable in the middle of the $4d$ 
  series which is linked to a large next-nearest neighbor antiferromagnetic exchange interaction. 
	We have calculated the energy dispersion of spin spirals for Ru, Mo, and Pd overlayers on Fe/W(001) using 
  a tungsten substrate consisting of eight atomic layers. Qualitatively, we obtain the same results as for 
  the corresponding quadlayers. Ru/Fe/W(001) exhibits a $p(2 \times 1)$ AFM ground
  state, but the DMI is significant and a local spin spiral minimum is only slightly higher in total energy. 
	For Mo we find an extremely small nearest-neighbor exchange interaction
  and a spin spiral ground state driven by competing exchange interactions. For Pd/Fe/W(001) the FM state is 
	the lowest in total energy amongst all considered magnetic configurations.
  However, the energy dispersion of spin spirals is very shallow and reminiscent to that of Fe/Ir(111) in which a 
	nanoskyrmion lattice has been discovered \cite{Heinze2011}.
  
  This paper is structured as follows. We begin with a description of the computational details and the methods which we used. In section III we discuss the results of our calculations.
  We start with the structural relaxations and the energy difference between the FM and $c(2 \times 2)$ AFM state. Then we present the spin spiral calculations for the $4d$/Fe/2W quadlayers.
  Finally, we present detailed studies for a Ru, Mo, and Pd overlayer on Fe/W(001) 
	including the effects of spin-orbit coupling, i.e.~DMI and magnetocrystalline anisotropy.
  
	 	\begin{figure}
 		\centering
    \includegraphics[width=0.99\linewidth]{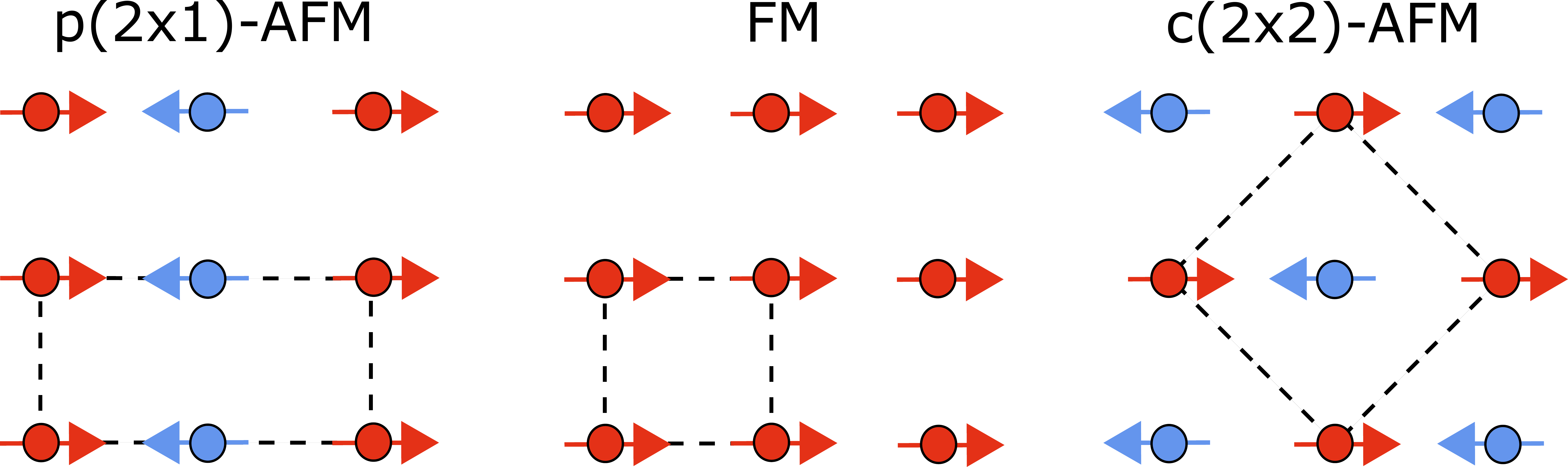}
 		\caption{The three collinear magnetic states with the corresponding two-dimensional unit cells. From left to right
    the $p(2\times 1)$ AFM, the FM, and the $c(2 \times 2)$ AFM state are shown.}
 		\label{fig:zeichnung}
 	\end{figure}
 			
 	\section{Computational details}\label{Kap: Methoden}
 	
 	We applied DFT as implemented in the full-potential linearized augmented plane wave method (FLAPW) \cite{Wimmer1981, Jansen1984} in film geometry \cite{Krakauer1979}.
  The structural, electronic, and magnetic properties of $4d$ transition-metal (TM) overlayers on Fe/W(001) were calculated using the J\"ulich DFT code \textsf{FLEUR}\cite{FLEUR}.
  The linearized augmented plane wave basis for the valence states was extended by local orbitals to 
	describe the 4\textit{s} and 4\textit{p} orbitals of the 4\textit{d} TM atoms and the 5\textit{p} 
	orbitals of the W atoms.  	
 	For all types of atoms we used a muffin-tin radius of 2.25~a.u. (1 a.u. = 0.529 \AA). The energy cutoff for the basis functions was \(k_{\rm max} = 4.1 \,\text{a.u.}^{-1}\) unless stated otherwise.  
  The experimental lattice constant of W was used (a$_{\text{W}} $~=~3.165~\AA) which is only by 0.5~\%
	smaller than the value obtained within the generalized gradient approximation of DFT \cite{Lazo2008}.
 	
	 	\begin{table}
 		\centering
 		\caption {Relaxed interlayer distances between the three topmost layers for $4d$ TM overlayers on Fe/W(001) in the FM state. 
    For comparison the distances of Fe/W(001) in the c$ (2\times 2) $ AFM state are given. All distances are given in a.u.}
 		\begin{ruledtabular}
 			\begin{tabular}{lccc}
 				         & \(d_{4\textit{d}-{\rm Fe}}\) & \(d_{{\rm Fe-W}}\) & \(d_{{\rm W}-{\rm W}}\) \\ \hline
 				Fe/W(001)    &                        &      2.44      &      2.97       \\
 				Nb/Fe/W(001) &          2.43          &      2.77      &      2.96       \\
 				Mo/Fe/W(001) &          2.21          &      2.70      &      2.97       \\
 				Tc/Fe/W(001) &          2.04          &      2.67      &      2.98       \\
 				Ru/Fe/W(001) &          2.04         &      2.65      &      2.94       \\
 				Rh/Fe/W(001) &          2.18         &      2.64      &      2.89       \\
 				Pd/Fe/W(001) &          2.53          &      2.43      &      2.90
 			\end{tabular}
 		\end{ruledtabular}
 		\label{Tab: Relaxation}
 	\end{table}
	
 	\subsection{Structural relaxations}\label{Kap: Methoden-Relaxation}
 	
    We used a symmetric film to calculate the relaxed interlayer distances for all 4\textit{d}/Fe/W(001) systems
		and for Fe/W(001). We considered the ferromagnetic (FM) and the c$ (2\times 2) $-antiferromagnetic (AFM) state. The substrate was represented by nine tungsten layers. An atomic layer of iron and an overlayer of the 
		4\textit{d} transition metal was added on both sides of the film.
    For the inner seven tungsten layers we fixed the interlayer distances to the experimental values. 
    All other interlayer distances were calculated by minimizing the forces to less than $10^{-5} $ hartree/a.u.~acting on the atoms \cite{Yu1991}.  
    We chose the general gradient approximation (GGA)\cite{Zhang1998} of the exchange-correlation potential. In the FM state, we used one atom per layer in the two-dimensional 
    unit cell and 840 \textit{k}-points in the full two-dimensional Brillouin zone (2D-BZ). For the c$(2\times 2) $ AFM state we used a two atomic two-dimensional unit cell 
    and 400 \textit{k}-points in the full 2D-BZ. 
    Since we found the FM state to be energetically more favorable than the c$ (2\times 2)$ AFM state for all $4d$ overlayers we have chosen the relaxed interlayer distances for the FM 
    state in all subsequent calculations (see Table \ref{Tab: Relaxation} for values). 
     	
 	\subsection{Spin-spiral calculations}\label{Kap: Methoden-spinmodell}
  
  We have calculated the energy dispersion $E({\mathbf q})$ of flat spin spirals \cite{Sandratskii1991,Kurz2004} 
  characterized by a vector ${\mathbf q}$ from the 2D-BZ. The magnetic moment ${\mathbf M}_i$ on lattice site 
	${\mathbf R}_i$ 
  is given for a flat spin spiral by $\mathbf{M}_i=M(\cos{(\mathbf{q} \mathbf{R}_i)},\sin{(\mathbf{q} \mathbf{R}_i)},0)$.
  We first performed self-consistent calculations without spin-orbit coupling applying the 
	generalized Bloch theorem \cite{Sandratskii1991,Kurz2004}. Based on these calculations we obtained the 
	energy contribution due to spin-orbit
  coupling, i.e.~from the DM interaction, for cycloidal spin spirals in first order perturbation theory as described in Ref.~\onlinecite{Heide2009,Zimmermann2014}.
  
  Spin spiral calculations were performed for freestanding 
  quadlayers consisting of a $4d$ TM layer, an Fe layer and two layers of the W(001) surface (denoted as $4d$/Fe/2W below)
  as well as for asymmetric films of a Mo, Ru, or Pd overlayer on an Fe layer and eight layers of the W(001) surface (denoted as $4d$/Fe/W(001) below). 
  For the later we have checked that increasing the thickness of the W substrate does not
  qualitatively change our results. The relaxed interlayer distances given in Table \ref{Tab: Relaxation} have been used.  
  For all freestanding quadlayers we used 1156 \textit{k}-points in the full 2D-BZ. 
  For Mo/Fe/W(001) and Ru/Fe/W(001) we also used 1156 \textit{k}-points and for Pd/Fe/W(001) 2304 \textit{k}-points were chosen in the full 2D-BZ.  
  Calculations were performed in local density approximation (LDA)~\cite{Vosko1980}. 
 
  The total energies of spin spiral calculations can be mapped 
	to the classical Heisenberg model on the two-dimensional atomic lattice of the Fe layer:
\begin{equation}\label{eq:ex}
H_{\rm ex}=-\sum_{ij}^{} J_{ij}\left(\textbf{m}_{i}\cdot \textbf{m}_{j}\right),
\end{equation}
where the exchange constants $J_{ij}$ denote the strength of the exchange interaction between the magnetic moments $\textbf{m}_{i}$ and $\textbf{m}_{j}$ located on
lattice sites $i$ and $j$. Here $\textbf{m}_{i}=\textbf{M}_{i}/{M}_{i}$ is the unit vector of the magnetic moment.
 Upon including spin-orbit coupling the Dzyaloshinskii-Moriya (DM) interaction arises
 \begin{equation}\label{eq-dm}
 H_{\rm DM}=-\sum_{ij}^{} \textbf{D}_{ij}\cdot \left(\textbf{m}_{i}\times\textbf{m}_j\right),
 \end{equation} 
 where $\textbf{D}_{ij}$ is the DM vector that denotes the strength and direction of the pairwise DM interaction between magnetic moments on the lattice.
The exchange constants $J_{ij}$ can be obtained
by fitting the energy dispersions of spin spirals calculated via DFT without spin-orbit coupling to the model given by 
Eq.~(1). The energy contribution of spin spirals due to spin-orbit coupling 
was used to determine the parameters $D_{ij}$ of the DM interaction. 
To describe the interaction of a given magnetic moment with the $i$-th shell of its nearest neighbours we introduce the shell resolved values $J_i$ and~$ \textbf{D}_i$.
The directions of the DM vectors are given by symmetry \cite{Dzyaloshinskii1957,Crepieux1998} (see e.g.~Ref.~\onlinecite{Meyer2017} for (001) bcc surface), while the magnitude and sign depend on the electronic structure.

	\begin{figure}
		\centering
    \includegraphics[width=0.99\linewidth]{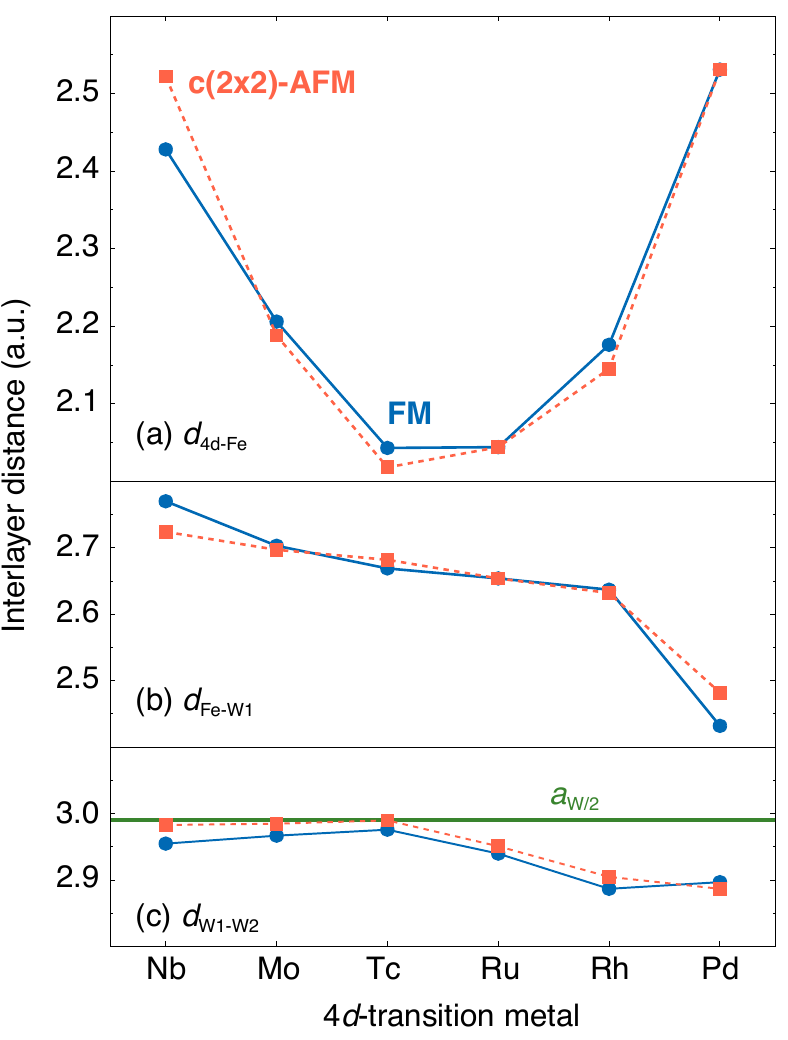}
		\caption{Relaxed interlayer distances between (a) the $4d$ overlayer and the Fe layer, $d_{4d\rm{-Fe}}$, (b) the Fe and the W surface layer, $d_{\rm{Fe-W}_1}$,
    and (c) the two topmost W layers, $d_{\rm{W}_1\rm{-W}_2}$, of the 4\textit{d}/Fe/W(001) films for the FM state 
		(blue filled circles)
    and for the c$(2 \times 2)$ AFM state (red filled circles). In (c) the unrelaxed interlayer distance of W bulk is 
		given as reference by the green line.
		}
		\label{fig:relaxation}
	\end{figure}

\subsection{Magnetocrystalline anisotropy}
The magnetocrystalline anisotropy energy (MAE), i.e.~the energy difference between a state with a magnetization perpendicular to the film and with an in-plane magnetization, has
been calculated for Mo/Fe/W(001), Ru/Fe/W(001), and Pd/Fe/W(001) applying the force theorem. We have used 
asymmetric films with 12 tungsten layers for the W(001) substrate in order to obtain converged
values of the MAE. First we performed a self-consistent scalar-relativistic calculation in LDA~\cite{Vosko1980} 
with $k_{\rm max} = 4.1~\text{a.u.}^{-1} $ and 1156~\textit{k}-points in the full 2D-BZ. 
Then we applied the force theorem \cite{Oswald1985,Liechtenstein1987} to evaluate the MAE. 
We 
performed calculations for a magnetization along the out-of-plane ($ \perp$) and the in-plane ($ \parallel $) direction 
based on the second variation method~\cite{Li1990}. Here, we used $k_{\rm max} = 4.3~\text{a.u.}^{-1}$ 
and 1936~\textit{k}-points. The obtained energy difference $K=E_{\perp} - E_{\parallel}$ can be included in the 
atomistic spin model by an uniaxial anisotropy term 
 \begin{equation}\label{eq-mae}
 H_{\rm MAE}=-\sum_{i}^{} K \left(m_i^z\right)^2.
 \end{equation} 

 	\section{Results}
 	
 	\subsection{Structural relaxations}
  
 	Figure \ref{fig:relaxation} shows the relaxed interlayer distances between the four uppermost layers of both sides of the symmetric $4d$/Fe/W(001) films. 
  Both the FM (orange curve) and the c$(2\times 2)$ AFM state (red curve) of the Fe layer have been considered. 
  Overall the differences in relaxations for these two magnetic configurations are small.	
  The interlayer distances between the 4\textit{d} transition-metal overlayer and the Fe layer show a parabolic curve 
  with respect to the $4d$ band filling. 
  Starting from Nb, the bonding orbitals are first filled until at the middle of the series antibonding states are also occupied.
	The distance between the Fe and the W layer [Fig.~\ref{fig:relaxation}~(b)] slightly decreases from Nb to Rh. 
	It is larger than the relaxed interlayer distance of 2.44 and 2.58~a.u.~reported
  for the FM and $c(2 \times 2)$ AFM state of Fe/W(001) \cite{Kubetzka2005} leading to a reduced 
	Fe-W hybridization in $4d$/Fe/W(001).
	A sharper drop is observed between Rh and Pd due to the complete filling of the $4d$ shell for Pd.   
	The variation of the distance between the upper two tungsten layers is small [Fig.~\ref{fig:relaxation}~(c)] and very close to the perfect unrelaxed value (green line with $a_{\rm W}/2$).  
	
	Since the differences of the structural relaxations between the FM and the c$ (2\times 2) $-AFM state are small, we performed all subsequent calculations only for the interlayer distances
  obtained for the FM state (cf.~Table \ref{Tab: Relaxation}) which is 
	energetically lower than the c$ (2\times 2) $-AFM for all $4d$ overlayers (cf.~section \ref{sec:FM_vs_AFM}). 
  For Fe/W(001) the c$(2\times 2)$-AFM state is the ground state and its interlayer distances were used.
  
	\subsection{FM vs.~$c(2\times 2)$ AFM state}
  \label{sec:FM_vs_AFM}
   	
		\begin{figure}
		\centering
		\includegraphics[width=0.99\linewidth]{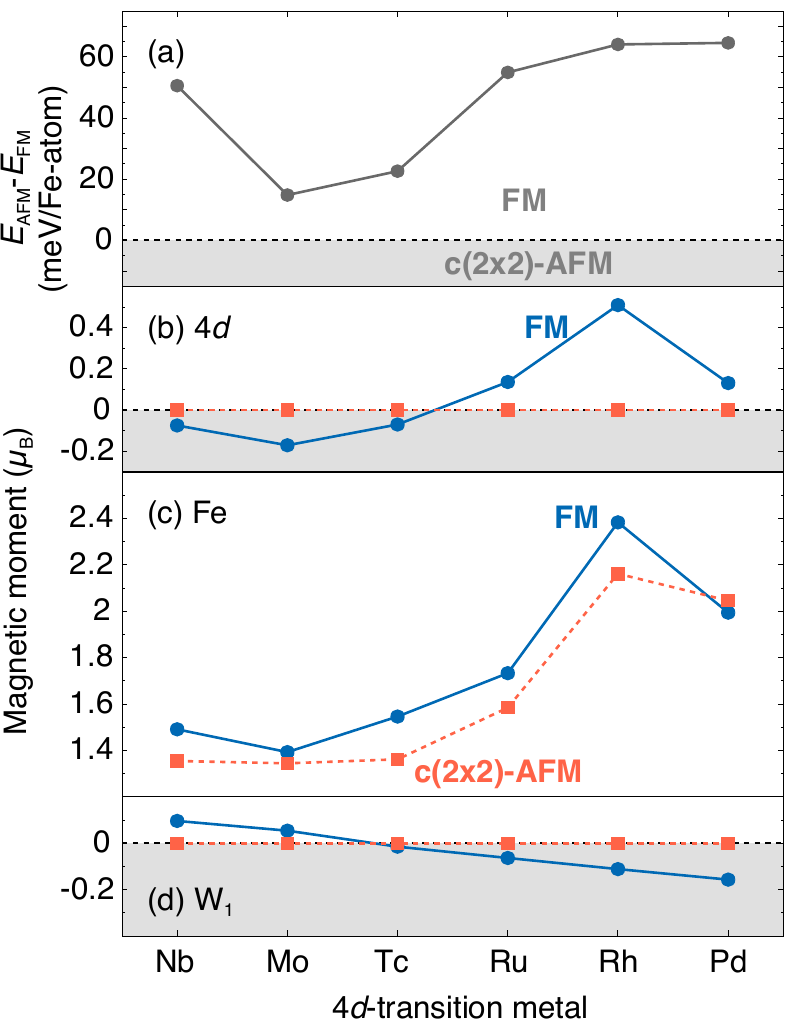}
		\caption{
			(a) Calculated energy differences $E_{\rm AFM}-E_{\rm FM}$ between the c$(2\times2)$-AFM and the FM state 
			for 4\textit{d}/Fe/W(001) films.
			In (b-d) the corresponding magnetic moments are shown for
			the upper three layers of the films, i.e.~the $4d$ overlayer,
			the Fe layer, and the upper W layer, in the FM (blue filled circles)
      and in the c$(2\times2)$-AFM state (red filled circles).
			}
		\label{fig:energie}
	\end{figure}

	\begin{figure*}
	\centering
  \includegraphics[width=0.99\linewidth]{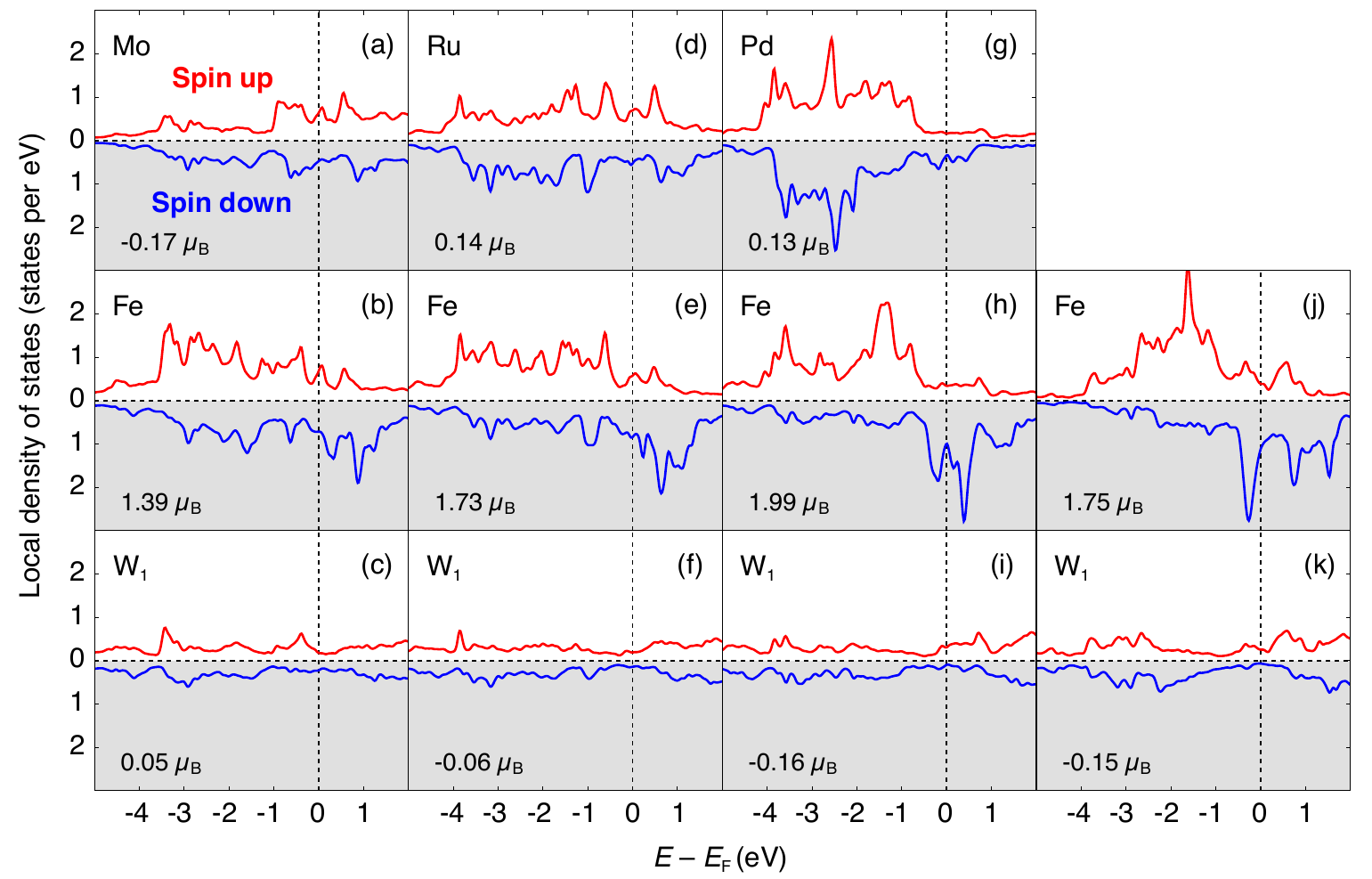}
	\caption{Spin-resolved local density of states in the ferromagnetic state for the topmost layers of (a-c) Mo/Fe/W(001), (d-f) Ru/Fe/W(001), (g-i) Pd/\-Fe/\-W(001),
  and (j-k) Fe/W(001). Red and blue curves show the majority and minority spin channel, respectively.	
	The magnetic moments of the different layers are given in the lower part of each panel.}
	\label{fig:dos}
\end{figure*}
		
 	Next we study how the 4\textit{d} transition-metal overlayer affects the energy difference between the FM and the c$ (2\times 2) $-AFM state
  for $4d$/Fe/W(001) films as shown in Fig.~\ref{fig:energie}~(a). 
  A positive sign of $\Delta E=E_{\rm AFM}-E_{\rm FM}$ indicates that the FM state is favorable while the AFM state is preferred for a negative value.
	If one restricts the Heisenberg model to nearest neighbors $\Delta E$ is directly proportional to $J_1$, i.e.~the
	exchange constant between nearest neighbor atoms.
	However, as we will see in the following sections in these systems exchange interactions beyond
	nearest neighbors need to be taken into account to describe the magnetic states.
	 	
	In agreement with previous studies \cite{Kubetzka2005} we find that Fe/W(001) prefers the c$ (2\times 2) $-AFM state by a large value of $125$~meV/Fe atom. 
	In contrast for all considered 4\textit{d}/Fe/W(001) films the FM state is lower in total energy than the 
	c$ (2\times 2) $-AFM state.
	However, the absolute value of the energy difference is relatively small.
	Hybridization with the 4\textit{d} transition-metal layer leads to lower energy differences, where Pd/Fe/W(001) has the largest energy difference with 
  65~meV/Fe atom and Mo/Fe/W(001) the smallest with 15~meV/Fe atom. 
  These results show that one can reduce the energy difference between the two magnetic states and thereby of the exchange interaction by an overlayer. 
  For the Pd overlayer this shift to a FM state 
  is consistent with the expectation from DFT calculations of $3d$ TM monolayers on Pd(001)~\cite{Blugel1988} which showed a FM ground state for Fe/Pd(001). 

  Since GGA tends to overestimate the AFM state the calculations shown in Fig.~\ref{fig:energie} have been performed 
	in LDA~\cite{Vosko1980} 
	using the relaxed interlayer distances obtained for the FM state (cf.~Table \ref{Tab: Relaxation}).
  Note, that we find qualitatively the same trend as a function of the $4d$ TM overlayer in GGA.
		
	The magnetic moments of the 4\textit{d}-transition metal, the Fe, and the W atoms of the topmost layers are shown in Fig.~\ref{fig:energie}~(b-d). 
	As expected, Fe has the largest magnetic moment. Its size depends significantly on the 4\textit{d}-transition metal due to hybridization between 
  the $3d$ and $4d$ states. At the beginning of the $4d$ series it is about 1.5~$\mu_{\rm B}$ in both magnetic states while it rises above 2~$\mu_{\rm B}$
  for the Rh and Pd overlayers. In the c$(2\times 2)$-AFM state the adjacent $4d$ and W layers obtain no induced magnetic moments due to symmetry.
  In the FM state the induced moments in the $4d$ layer change from negative to positive with $4d$ band filling due to the change of their spin susceptibility.
  For Rh the largest magnetic moment of about 0.4~$\mu_{\rm B}$ is found.
  The induced magnetic moment of the W atoms at the interface is opposite to that of the corresponding 4\textit{d} atoms. 
	
	To obtain more insight into the influence of the $4d$ overlayer on the electronic and magnetic properties of the films, 
  we present in Fig.~\ref{fig:dos} the local density of states (LDOS) of the topmost three layers for the Mo, Ru, and Pd overlayer on Fe/W(001). 
	Compared to Fe/W(001) in all 4\textit{d}/Fe/W(001) systems the LDOS of Fe becomes broader and flatter due to the increased coordination and
  strong hybridization of $3d$ and $4d$ states. This leads to the strongly reduced exchange splitting and magnetic moment at the beginning of the $4d$ series.
	The shape of the LDOS for all the 4\textit{d} transition-metal overlayers is similar and shows the filling 
	of the $4d$ band.
	The LDOS narrows and increases due to the reduction of the extent of the $4d$ orbitals from Nb to Pd.
	The hybridization with the Fe $3d$ states induces a small magnetic moment in the $4d$ overlayer.
	One can see a number of peaks in the vicinity of the Fermi energy which appear in both the overlayer 
	and the Fe layer, prominently in the majority spin channel for Mo and Fe (Figs.~\ref{fig:dos}(a,b))
	and Ru and Fe (Figs.~\ref{fig:dos}(d,e)), indicating the pronounced $3d$-$4d$ hybridization.
  The LDOS of the W surface layer is rather flat due to the larger extent of the $5d$ states resulting in
	a larger band width. The LDOS of the W layer also displays a hybridization with the Fe layer which shows
	most clearly in the absence of an overlayer, i.e.~for Fe/W(001) (Figs.~\ref{fig:dos}(j,k)).
	Some of the hybrid states show peaks in the $4d$, Fe, and W LDOS which is therefore also affected by
	the $4d$ overlayer.
	
 	\subsection{Spin spirals in unsupported quadlayers}
  
  So far, we have studied only two collinear magnetic states, the FM and the c$(2 \times 2)$ AFM state. To understand the effect of the hybridization
  with the $4d$ overlayer on the exchange interactions in the Fe layer we need to expand our investigation to non-collinear magnetic states such as spin spirals.
  From their energy dispersion we can obtain the exchange constants as discussed in the method section
\ref{Kap: Methoden}. However, spin spiral calculations are computationally
  very demanding and time consuming. Therefore, we focus in this section on model systems consisting of only four layers: the $4d$ overlayer, the Fe layer, and
  two layers of the W(001) substrate which we denote as unsupported quadlayers $4d$/Fe/2W. Our studies show that two W layers are already sufficient to obtain
  qualitatively the correct trends concerning the magnetic ground state as long as we neglect spin-orbit coupling. To obtain even a qualitatively reasonable 
  description of the Dzyaloshinskii-Moriya interaction or the magnetocrystalline anisotropy more W layers are needed. Film calculations for selected overlayers
  are discussed in section \ref{sec:films}.
 	 	
 	\begin{figure}
	\centering
  \includegraphics[width=0.99\linewidth]{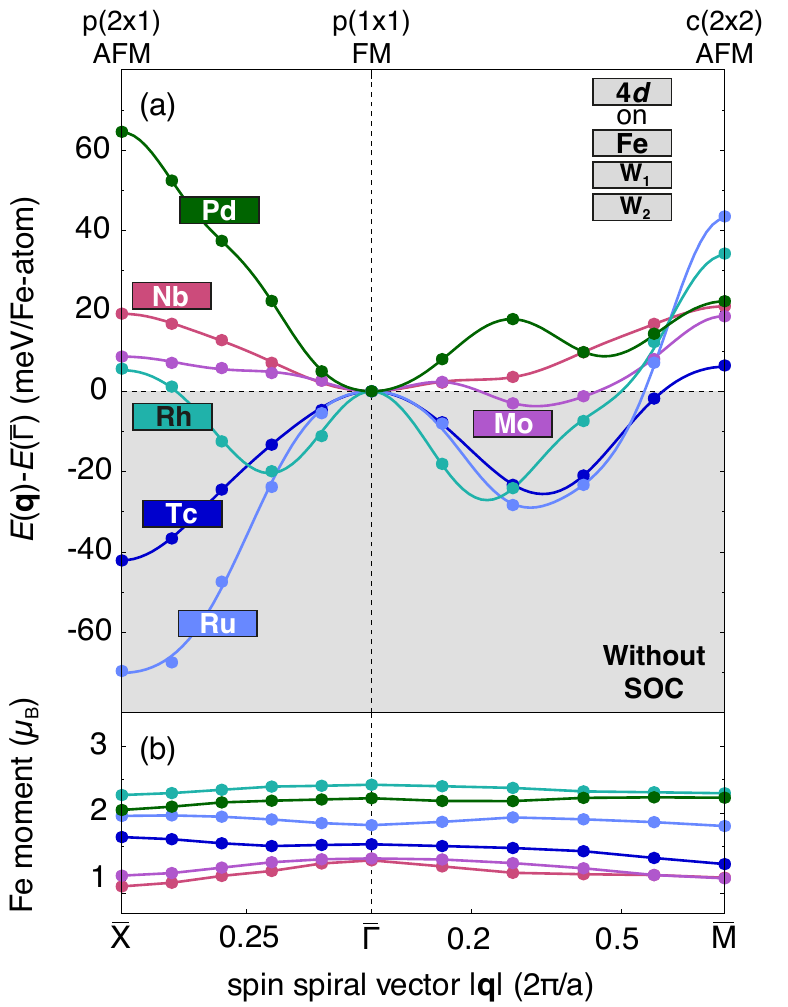}
	\caption{(a) Energy dispersion $E({\mathbf q})$ of spin spirals without spin-orbit coupling
  along the high symmetry directions for unsupported $4d$/Fe/2W quadlayers. The filled symbols 
  represent the points obtained from DFT and the lines are fits to the Heisenberg model. The
  collinear magnetic states at the three high symmetry points are given at the top of the graph.
  (b) Magnetic moments of the Fe layer as a function of the spin spiral vector.
  }
	\label{fig:qud_ss}
\end{figure}

\begin{figure}
	\centering
  \includegraphics[width=0.99\linewidth]{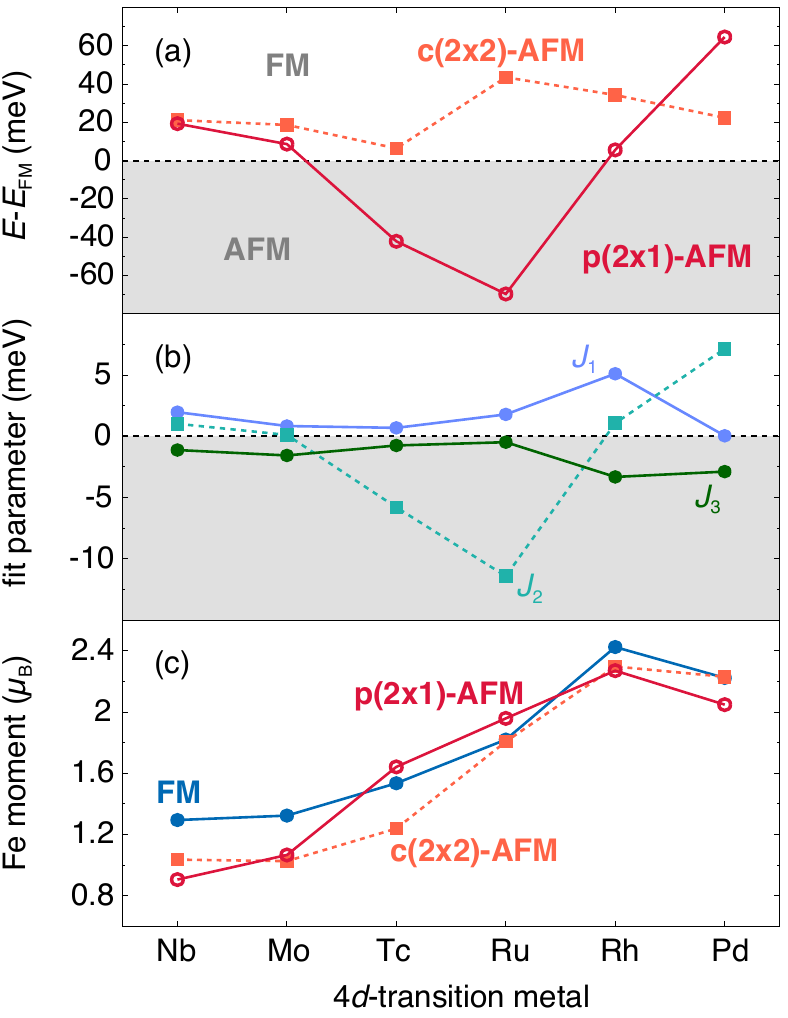}
	\caption{(a) Energy of the p$ (2\times 1) $- and the c$ (2\times 2) $-AFM states with respect to the 
	FM state for $4d$/Fe/2W unsupported quadlayers,
  i.e.~obtained from the energy dispersion of spin spirals shown in Fig.~\ref{fig:qud_ss}(a).
  A positive (negative) sign denotes that the FM (AFM) state is more favorable.
  (b) First three exchange constants obtained from fitting the energy dispersion of spin spirals and
  (c) magnetic moments of the Fe layer in the three collinear magnetic states at the high symmetry
  points of the Brillouin zone (cf.~Fig.~\ref{fig:qud_ss}(a)).
}
	\label{fig:quad_ausw}
\end{figure}
	
  The calculated energy dispersion $E({\mathbf q})$ of spin spirals for the quadlayers are shown in 
	Fig.~\ref{fig:qud_ss}(a). The solid
  circles are the energies obtained from DFT without SOC
	and the lines are fits to the Heisenberg model, Eq.~(\ref{eq:ex}).
  At the high symmetry points collinear states are obtained: the $\bar{\Gamma}$ point (${\mathbf q}=0$) corresponds to the FM state, the
  $\bar{\rm M}$ point to the c$(2 \times 2)$ AFM state, and the $\bar{\rm X}$ point to the p$(2 \times 1)$ AFM state
	(cf.~Fig.~\ref{fig:zeichnung}).
	
	As a general trend we note from Fig.~\ref{fig:qud_ss}(a) that as the $4d$ overlayer is varied the energy of the 
	p$(2 \times 1)$-AFM state first decreases from Nb 
  ($\approx +20$~meV) to Ru ($\approx -70$~meV) and then rises again up to a value of $\approx +65$~meV for Pd. 
	The energy of the c$(2 \times 2)$-AFM state, on the
  other hand, shows no change of sign in qualitative agreement with the observation from the film calculations 
	presented in Fig.~\ref{fig:energie}. These 
  trends for the energy difference between the collinear magnetic states are summarized for the $4d$/Fe/2W quadlayers 
	in Fig.~\ref{fig:quad_ausw}~(a). Quantitative
  discrepancies with the film calculations are not surprising since the thickness of the substrate certainly influences the result of the calculations. Importantly,
  we find that it is the p$(2 \times 1)$-AFM state which is most influenced by the hybridization between Fe and the $4d$ overlayer. 
	
 	The shape of the dispersion curves $E({\mathbf q})$ as well as the global energy minima 
	(summarized in Table~\ref{tab-minima_ql}) 
	strongly varies with the 4\textit{d} transition-metal overlayer and its band filling.
  For Nb and Mo overlayers we observe extremely flat energy dispersion curves compared to all other quadlayers. 
  In Nb/Fe/2W the energy minimum is at the FM state, while 
  a spin spiral with a period of $ \lambda = 0.96 $~nm along $\overline{\Gamma {\rm M}}$ is the lowest state for Mo/Fe/2W.
	Interestingly, there is no minimum in the other high symmetry direction $\overline{\Gamma {\rm X}}$ for Mo/Fe/2W.
 	This suggests a strong directional anisotropy of spin spirals in Mo/Fe/W(001).
  
	The energy dispersion for Tc/Fe/2W and Ru/Fe/2W exhibit a larger energy scale and the FM and c$(2 \times 2)$ AFM
	are unfavorable. The global energy minimum for these two systems is at the p$(2 \times 1)$-AFM state.
  In addition, there is a local minimum for a spin spiral along the $ \overline{\Gamma {\rm M}}$ direction
	with a period of about 0.9~nm (Tc) and 1~nm (Ru).
  For the Ru overlayer the minimum at the p$ (2\times 1)$-AFM state is much lower than for Tc indicating a 
	stronger antiferromagnetic exchange interaction.

 	Rh/Fe/2W has two spin spiral minima with nearly the same energy. One minimum lies in $ \overline{\Gamma {\rm X}}$- 
	and the other one in $ \overline{\Gamma {\rm M}}$-direction, the latter being energetically slightly deeper. 
 	The spin spirals have periods of $ \lambda_{\overline{\Gamma {\rm X}}} = 1.62 $~nm and 
	$\lambda_{\overline{\Gamma {\rm M}}} = 1.41 $~nm. Finally, in Pd/Fe/2W the global 
  minimum is at the FM state as at the beginning of the $4d$ series for Nb. However, the energy rises much more 
	quickly in the vicinity of the $\overline{\Gamma}$ point of the Brillouin zone. 
  
  The magnetic moments of the Fe layer in the unsupported quadlayers [Fig.~\ref{fig:qud_ss}(b)] show a similar rise with increased band filling of the $4d$ 
  overlayer as observed previously in the film calculations [Fig.~\ref{fig:energie}(c)]. However, as a function of spin spiral vector (or period) we find only
  a relatively small variation of the moment. This underlines that a fit to the Heisenberg model, which rests on the assumption of constant magnetic moments,
  to extract exchange constant is reasonable.
  
  From the energy dispersion $E({\mathbf q})$ of spin spirals [Fig.~\ref{fig:qud_ss}(a)] 
  we have obtained the exchange interactions by fitting to the Heisenberg model, Eq.~(\ref{eq:ex}). The values of the first three 
  exchange constants are shown in Fig.~\ref{fig:quad_ausw}(b). The behavior of the energy differences between the collinear states is also reflected in the 
  exchange constants.
  The curve showing $J_1$ is qualitatively analogous to the energy difference between the FM and the 
	c$ (2 \times2) $-AFM state. 
	It indicates that all systems are characterized by a small ferromagnetic nearest-neighbor exchange interaction. 
	For comparison, we note that $J_1 \approx +22$~meV for an unsupported Fe monolayer on the W(001) lattice constant
	and $J_1=-26$~meV for Fe/W(001) \cite{Ferriani2007}. Interestingly, we do not find a change of sign of $J_1$
	with band filling of the overlayer as has been reported previously for Fe monolayers on transition-metal 
	surfaces \cite{Ferriani2007,Hardrat2009}.
	
 	The trend of $J_2$ follows the energy of the p$(2\times 1)$ AFM state, i.e.~it is "v"-shaped 
	[cf.~Fig.~\ref{fig:quad_ausw}(a)] and becomes strongly antiferromagnetic in the center of the $4d$ series. 
  For Nb, Mo, Rh, and Pd overlayers the exchange constants for the first three neighbors are of similar magnitude 
	but vary in sign characteristic of exchange frustration. 
	For $J_3<0$ this can lead to spin spiral ground states (cf.~phase diagrams shown in Ref.~\onlinecite{Ferriani2007})
	as observed for Rh/Fe/2W and Mo/Fe/2W.
  For quadlayers with Tc and Ru the exchange constant between second nearest 
	neighbors, $J_{2}$, is negative and dominates resulting in the collinear 
  p$(2\times 1)$ AFM state being lowest.
	
	The magnetic moments of the Fe layer rise in the quadlayers from low values of about $1 \mu_{rm B}$ for a Nb overlayer
	to about $2.4 \mu_{\rm B}$ for the Rh overlayer [Fig.\ref{fig:quad_ausw}(c)]. The trend with the $4d$ overlayer as well
	as the order of magnitude of the magnetic moments is very similar to that observed for the Fe layer in the film systems 
	discussed in Fig.~\ref{fig:energie}(c).
 	
 	 	\begin{table}
 		\centering
 		\begin{ruledtabular}
 		\begin{tabular}{lcc}
 	 			     & \textbf{q} in $ 2\pi/a $ & magnetic state \\ \hline
 			Nb/Fe/2W &                  (0.00,0.00)                   &                 FM           \\
 			Mo/Fe/2W &                       (0.23,0.23)                   &           spin spiral ($ \lambda = 0.96   $~nm)                \\
 			Tc/Fe/2W &                      (0.50,0.00)                      &            p$ (2\times 1) $-AFM             \\
 			Ru/Fe/2W &                 (0.50,0.00)                  &                     p$ (2\times 1) $-AFM    \\
 			Rh/Fe/2W &                     (0.16,0.16)                     &              spin spiral ($ \lambda = 1.41    $~nm)                \\
 			Pd/Fe/2W &                  (0.00,0.00)                   &                    FM         \\ 
 		\end{tabular} 
 	\end{ruledtabular}
 		\caption{Global energy minima extracted from the energy dispersion of spin spirals for unsupported 
		4\textit{d}/Fe/2W quadlayers. 
    The spin spiral vector \textbf{q} of the minimum and the associated magnetic state are given.}
 		\label{tab-minima_ql}
 	\end{table}
 	
  Based on the results presented in this section and in the previous section we conclude that one can significantly tune the energy difference between the FM state and the 
  c$(2 \times 2)$-AFM state via hybridization with a 4\textit{d} transition-metal overlayer as discussed in the introduction. However, the spin spiral 
  calculations for quadlayers demonstrate that the variation of the energy difference between the FM state and the 
	$p(2 \times 1)$-AFM state is influenced
  much more strongly, even displaying a transition, which is reflected in the exchange constant between second nearest    
	neighbors becoming negative and dominant in the middle of the $4d$ series.
  
It is remarkable that we can obtain quite a variety of magnetic properties of the Fe layer in terms of the global energy minima, the magnetic moments, and the exchange constants. Due to the strong frustration in the exchange interaction, the systems with Nb, Mo, Rh, and Pd overlayers are promising candidates for the stabilization of complex non-collinear magnetic states, so further investigation in a more realistic film structure
is worthwhile. The system with Ru is also interesting in this context, although the frustration is smaller. 
However, this system can be considered as a candidate for the stabilization of skyrmions in a 
p$(2\times1)$-AFM background \cite{Menezes2020}.
 	 	
 	\subsection{Film calculations}\label{sec:films} 	
 	
  	 	All 4\textit{d}/Fe/W(001) systems seem to be suitable candidates for the stabilization of 
			non-collinear magnetic states based on the 
      results within the approximation of quadlayers presented in the previous section.
      Here we focus on three representative systems
      for a detailed study: Mo/Fe/W(001), Ru/Fe/W(001) and Pd/Fe/W(001). 
			
			Note, that we have not
			studied Rh/Fe/W(001) since our key interest lies in finding spin spiral minima in the vicinity
			of the antiferromagnetic states or with magnetic properties distinctively different from other
			film systems studied previously. The results on Rh/Fe/2W
			quadlayers strongly suggest that the corresponding film system will exhibit small period spin spirals
			stabilized by frustrated exchange interactions. In this respect, this system is similar to
			fcc-Rh/Fe/Ir(111) for which such a spin spiral ground state has been observed by spin-polarized
			scanning tunneling microscopy experiments \cite{Romming2018}.
      
      As we will see below the system with the Mo overlayer displays the smallest value of the
      nearest-neighbor ferromagnetic exchange constant. Due to exchange beyond nearest neighbors a spin spiral ground state is stabilized. For 
      the Ru overlayer the nearest-neighbor ferromagnetic exchange is enhanced, however, a row-wise antiferromagnetic ground state is obtained
      due to a much larger antiferromagnetic next-nearest neighbor exchange. Finally, the nearest-neighbor ferromagnetic exchange dominates for
      the Pd overlayer system and we find the ferromagnetic state to be lowest among all considered states. However, exchange beyond nearest-neighbors 
      is significant and higher-order exchange interactions may lead to a non-collinear spin structure. In none of the systems, the Dzyaloshinskii-Moriya 
      interaction is large enough to enforce a non-collinear ground state. 
      
\begin{figure}
 	 		\centering
      \includegraphics[width=0.99\linewidth]{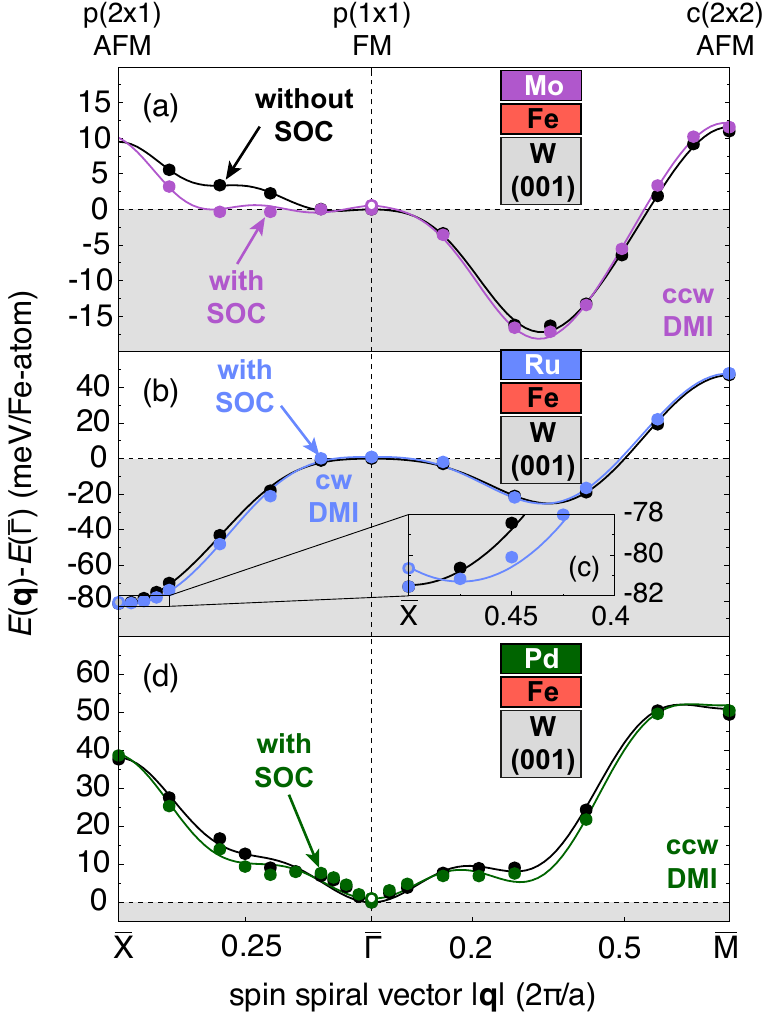}
 	 		\caption{ Energy dispersion $E({\mathbf q})$ of spin spirals without (black) and with the contributions of spin-orbit coupling (orange, red) composed of the DM interaction and the magnetocrystalline anisotropy, for (a) Mo/Fe/W(001), (b,c) Ru/Fe/W(001) and (d) Pd/Fe/W(001).	The circles represent the data points calculated
      via DFT and the lines are fits to the atomistic spin model. In Mo/Fe/W(001) and Pd/Fe/W(001) a counter-clockwise (ccw) and in Ru/Fe/W(001) a clockwise (cw) rotation sense is preferred. In all four panels the influence of the magnetocrystalline anisotropy is indicated by an open circle. }
 	 		\label{fig:film_ss}
 	 	\end{figure}

      The calculated energy dispersions of spin spirals without and with spin-orbit coupling         
      for the three film systems are displayed in Fig.~\ref{fig:film_ss}. The magnetocrystalline anisotropy energy has been calculated 
      for the collinear magnetic state with the lowest total energy.       
      We discuss the systems in detail below one by one.
  	 	
	\subsubsection{Mo/Fe/W(001)}\label{Kap: Mo}

From the quadlayer calculations we anticipate for Mo/Fe/W(001) that the energy scale of spin spirals
is very small and that the energy landscape is anisotropic (cf.~Fig.~\ref{fig:qud_ss}(a)). The result for the film calculation with a W(001) substrate with 8 layers
shown in Fig.~\ref{fig:film_ss}~(a) is qualitatively in agreement with this expectation. We observe a deep spin spiral energy minimum along the $\overline{\Gamma {\rm M}}$ direction with a short period of $\lambda = 0.90$~nm, similar to that of the quadlayer (cf.~Tab.~\ref{tab-minima_ql}), 
while the dispersion is almost flat along the other high symmetry direction. 
The Fe magnetic moment is about $1.4 \mu_{\rm B}$ and varies only little as a function of spin spiral vector (not shown)
as in the quadlayer calculations.
This provides further evidence that the quadlayers are a good approximation
to obtain the general trends concerning the change of the exchange interaction due to the $4d$ overlayer. 

 	 	\begin{table*}
 	 		\centering
 	 		\caption {Calculated magnetocrystalline anisotropy energy for $4d$ overlayers on Fe/W(001). For each system the value of $K$ is given in meV
      per Fe atom, the easy magnetization axis, and the corresponding magnetic state for which the calculation was performed. As a reference the
      value for Fe/W(001) is given from Ref.~\onlinecite{Kubetzka2005}.}
 	 		\begin{ruledtabular}
 	 			\begin{tabular}{lccc}
 	 				                            						 & $K$ &    preferred direction          &    magnetic state    \\ \hline
 	 				Mo/Fe/W(001)                 &       $-1.1$        &   in-plane   &          FM          \\
 	 				Ru/Fe/W(001)                 &       $-1.8$        &   in-plane   & p$ (2\times1) $-AFM  \\        
 	 				Pd/Fe/W(001)                 &       2.1        & out-of-plane &          FM \\
          Fe/W(001)\cite{Kubetzka2005} &    2.4               & out-of-plane & c$ (2\times 2) $-AFM 
 	 			\end{tabular}
 	 		\end{ruledtabular}
 	 		\label{Tab:anisotropie}
 	 	\end{table*}

	\begin{table*}
	\centering
	\caption {Exchange constants $J_i$ for $i$-th nearest neighbors obtained from fits to the energy dispersion of spin spirals without
  spin-orbit coupling obtained from DFT calculations. All values are given in meV.}
	\begin{ruledtabular}
		\begin{tabular}{lccccccccc}
			meV          & $ J_1 $ & $ J_2 $ & $ J_3 $ & $ J_4 $ & $ J_5 $ & $ J_6 $ & $ J_7 $ & $ J_8 $ & $ J_9 $ \\\hline
			Mo/Fe/W(001) &  0.92   &  0.43   &  $-3.09$  & 0.06    & 1.17    & 0.14    & 0.02    & 0.13    & $-0.43$   \\
			Ru/Fe/W(001) &  3.60   &  $-13.11$   &  0.47   &  0.85       &    0.44     &   0.57      &     &         &  \\
			Pd/Fe/W(001) &  6.44   &  $-0.42$   &  $-2.48$   &   0.10      &   0.41      &   $ - $0.18      &  1.01       
			& $-0.06$       &
		\end{tabular}
	\end{ruledtabular}
	\label{tab:J}
	\end{table*}

	\begin{table*}
	\centering
	\caption {Dzyaloshinskii-Moriya interaction constants $D_i$ for $i$-th nearest neighbors obtained from fits to the energy dispersion of spin spirals
  including spin-orbit coupling obtained from DFT calculations. All values are given in meV.}
	\begin{ruledtabular}
		\begin{tabular}{lccccc}
			meV          & $ D_1 $ & $ D_2 $ & $ D_3 $ & $ D_4 $ & $ D_5 $ \\\hline
			Mo/Fe/W(001) &  $ - $0.40  &  $ - $0.97  &  0.40   &  0.10   &         \\
			Ru/Fe/W(001) &  0.18   &  1.44   &  $ - $0.51  &  $ - $0.13  &  $ - $0.35  \\
			Pd/Fe/W(001) &  $ - $1.38  &  $ - $0.34  &  0.31   &  0.17   &  $ - $0.01   		\end{tabular}
	\end{ruledtabular}
	\label{tab:D}
	\end{table*}

Upon including spin-orbit coupling the energetic degeneracy between clockwise (cw) and counterclockwise (ccw) rotating spin spirals is lifted due to the
Dzyaloshinskii-Moriya interaction. By symmetry of the DMI
cycloidal spin spirals are preferred and for Mo/Fe/W(001) the counterclockwise rotational sense is 
favorable (denoted as ccw DMI in Fig.~\ref{fig:film_ss}(a)). The spin spiral energy minimum becomes slightly 
deeper due to the effect of the DMI.

Spin-orbit coupling further results in the magnetocrystalline anisotropy which leads to a preference of collinear magnetic states over spin spiral states.
The magnetocrystalline anisotropy energy, i.e.~the energy difference between an out-of-plane and an in-plane magnetization direction, denoted as $K$ 
shifts the total energy of spin spirals by $K/2$ in the limit of large spin spiral periods, i.e.
\begin{equation}\label{key}
\left|\frac{E_{{\rm MAE}}}{N}\right|  = \left|-\frac{1}{N}\sum_{i=1}^{N}  K \, \left(\hat{e}_z\cdot\vec{S}_i\right)^2\right|\approx\left|\frac{K}{2}\right|.
\end{equation}
$N$ is the number of magnetic moments averaged over and $ \hat{e} _z $ a direction vector that is perpendicular to the surface. We have calculated the
magnetocrystalline anisotropy energy in the FM state which is the energetically lowest among the collinear magnetic states.
Note that the magnetocrystalline anisotropy energy depends on the electronic structure and therefore it can be sensitive to the considered magnetic state.
For Mo/Fe/W(001) we obtain an in-plane anisotropy with $ K = 1.1 $~meV per Fe atom. The shift of the spin spirals with respect to the FM state 
is marked by an open circle at the $\overline{\Gamma}$ point (FM state) in Fig.~\ref{fig:film_ss}~(a). 

The calculated energy dispersion $ E (\textbf{q}) $ of flat spin spirals is used to determine the parameters of the atomistic spin model. The corresponding fits are shown as lines in Fig.~\ref{fig:film_ss}. The energies without 
spin-orbit coupling, are used to find the parameters of the exchange constants (Tab.~\ref{tab:J}). Nine shells of the nearest neighbors are considered. The energy contribution to spin spirals due to spin-orbit coupling 
$\Delta E_{\rm DMI}(\textbf{q})$ is used to determine the parameters of the DM interaction. For these it is sufficient to consider four nearest neighbor shells to obtain a good fit (Tab.~\ref{tab:D}). 

The exchange interaction is frustrated (Tab.~\ref{tab:J}), i.e.~a positive value of the nearest- and next-nearest neighbor exchange $J_1$ and $J_2$ preferring ferromagnetic alignment compete with a much larger negative value of 
$J_3$ which favors an antiferromagnetic alignment. 
The relative strengths and signs of the exchange constants
are consistent with those obtained from the quadlayer calculations (cf.~Fig.~\ref{fig:quad_ausw}(b)).
The absolute values of all exchange constants
are very small compared to the nearest-neighbor exchange constant $J_1=-26$~meV reported for 
Fe/W(001) \cite{Ferriani2007}. 
The DM interaction also shows considerable frustration (Tab.~\ref{tab:D}). The largest contribution comes from the 
second shell of the neighbors $ D_2 <0 $. However, there are quite large competing contributions from the first and 
third shell of neighbors.

Our results for Mo/Fe/W(001) demonstrate how drastically the magnetic interactions in the Fe monolayer can be modified due to the hybridization with a $4d$ overlayer. An experimental study therefore seems worthwhile. 
Given the strong exchange frustration and the small absolute values of the exchange constants, it is likely that 
higher-order exchange interactions such as the 4-spin or the biquadratic interaction could lead to the stabilization 
of a two-dimensional non-collinear magnetic state.

 	\subsubsection{Ru/Fe/W(001)}\label{Kap:Ru}
  The spin spiral energy dispersion of Ru/Fe/W(001) [Fig.~\ref{fig:film_ss}~(b)] shows that the ground state is the 
	p$ (2\times 1) $ row-wise antiferromagnetic state in agreement with the expectation from the quadlayer 
	calculations (cf.~Fig.~\ref{fig:qud_ss}).
	We obtain a magnetic moment for Fe of about $1.7 \mu_B$ at the FM state which varies with 
	the spin spiral vector as observed in the quadlayer.

	 We map the total energy calculations from DFT to the atomistic spin model to obtain further insight into the magnetic 
	interactions. For the exchange interaction we need to take six shells of neighbors into account (Tab.~\ref{tab:J}). 
	The exchange interaction is dominated by $ J_2 \approx -13$~meV, which favors an antiferromagnetic alignment of next-nearest-neighbor magnetic moments and leads to the p$ (2\times 1) $ row-wise antiferromagnetic ground state.
	This term has to compete with the ferromagnetic nearest-neighbor exchange 
    $ J_1=3.6$~meV which is already much larger than for Mo/Fe/W(001), however, still extremely small compared to 
		that of e.g.~a freestanding Fe monolayer on the tungsten lattice constant ($J_1 \approx 22$~meV). We also observe 
		frustration of the DMI as seen from the values given in Table~\ref{tab:D}.

  Due to the large in-plane magnetocrystalline anisotropy of $ K = 1.8 $~meV 
	the	p$ (2\times 1) $ row-wise antiferromagnetic state
	is also favorable with respect to a local spin spiral minimum at 
  $  q = 0.477 \cdot 2\pi/a $ ($ \lambda \approx 13.8$~nm). 
	However, the spin spiral which gains energy due to the DMI is in total only $ 0.5 $~meV/Fe atom 
  higher in energy as seen from the inset Fig.~\ref{fig:film_ss}~(c). 
  Therefore, an experimental study would be very interesting.
	
	In principle, higher-order exchange interactions \cite{Hoffmann2020} could also stabilize a more
  complex non-collinear state at the superposition of two spin spirals at the two 
	$\overline{\rm M}$ points of the 2D-BZ a so-called 2Q state \cite{Ferriani2007}. 
	However, we have calculated the 2Q state for Ru/Fe/W(001) and found that it is 
	energetically unfavorable by 14~meV/Fe-atom with respect to the row-wise AFM state.
  
 	\subsubsection{Pd/Fe/W(001)}\label{Kap:Pd}
 	
  For Pd/Fe/W(001) we conclude from the energy dispersion in Fig.~\ref{fig:film_ss}~(d)
  that the ferromagnetic state ($\overline{\Gamma}$ point) is lowest among all spin spiral states even upon including spin-orbit coupling. For the ferromagnetic
  state the magnetocrystalline anisotropy energy of $ K = 2.1 $~meV/Fe atom is quite strong and 
	favors an out-of-plane magnetization direction. The energy contribution due to DMI 
  is relatively small for spin spiral states in the vicinity of the $\overline{\Gamma}$-point. 
	However, the overall energy scale of the energy dispersion is small
  compared to an unsupported Fe ML indicating the strong influence of the hybridization of the Fe layer with the adjacent    
	Pd and W layers. The energy dispersion
  is surprisingly very similar to that of an Fe ML on Ir(111) which exhibits a nanoskyrmion lattice \cite{Heinze2011}. 
  
	The atomistic spin model is used to describe the interactions. It turns out that both the energy dispersion without and with spin-orbit coupling around the FM state ($\overline{\Gamma}$-point) has a curvature that can not be perfectly described by the model. Even increasing the number of adjacent neighbors to as many as thirteen shells does not lead to a significant improvement. However, the energy scale is extremely small.
	As best solution for the exchange interaction 
	seven shells of the nearest neighbors are taken into account (Tab.~\ref{tab:J}) and for the DM interaction five shells 
	(Tab.~\ref{tab:D}). 

	Compared to Mo/Fe/W(001) and Ru/Fe/W(001) the nearest-neighbor ferromagnetic exchange further increases to
	$J_1 \approx 6.5$~meV and is the dominating exchange constant (see Tab.~\ref{tab:J}). 
  However, $J_1$ is still small enough such that exchange beyond nearest-neighbors can compete which is reflected in 
	the complex shape of the energy
  dispersion curve (Fig.~\ref{fig:film_ss}~(d)). In particular, $J_3 \approx -2.5$~meV is of a similar order of magnitude. 
	The DMI is similarly dominated by
  the nearest-neighbor interaction, but there is also some competition with terms beyond nearest neighbors. 
  
As expected from the similar energy dispersion curves the exchange constants of Pd/Fe/W(001) are quite similar to those obtained for an Fe monolayer on Ir(111) 
($ J_1 = 5.7 $~meV, $ J_2 = -0.84 $~meV, $ J_3=-1.45 $~meV, values from Ref.~\onlinecite{Heinze2011}). Fe/Ir(111) also possesses an out-of-plane magnetocrystalline anisotropy, albeit with a significantly smaller value of
$K=0.8$ meV/Fe atom, and a strong DM interaction of $D=1.8$ meV.
Although the different lattice structure -- square vs.~hexagonal --
makes a one-to-one comparison difficult it is still intriguing to see the similarities which suggest that 
Pd/Fe/W(001) could also exhibit a complex magnetic ground state.
Therefore, an experimental determination of the ground state of this film system appears to be worthwhile.
 	
  \subsubsection{Extrapolated exchange energy landscape}
  
	\begin{figure}
	\centering
  \includegraphics[width=0.99\linewidth]{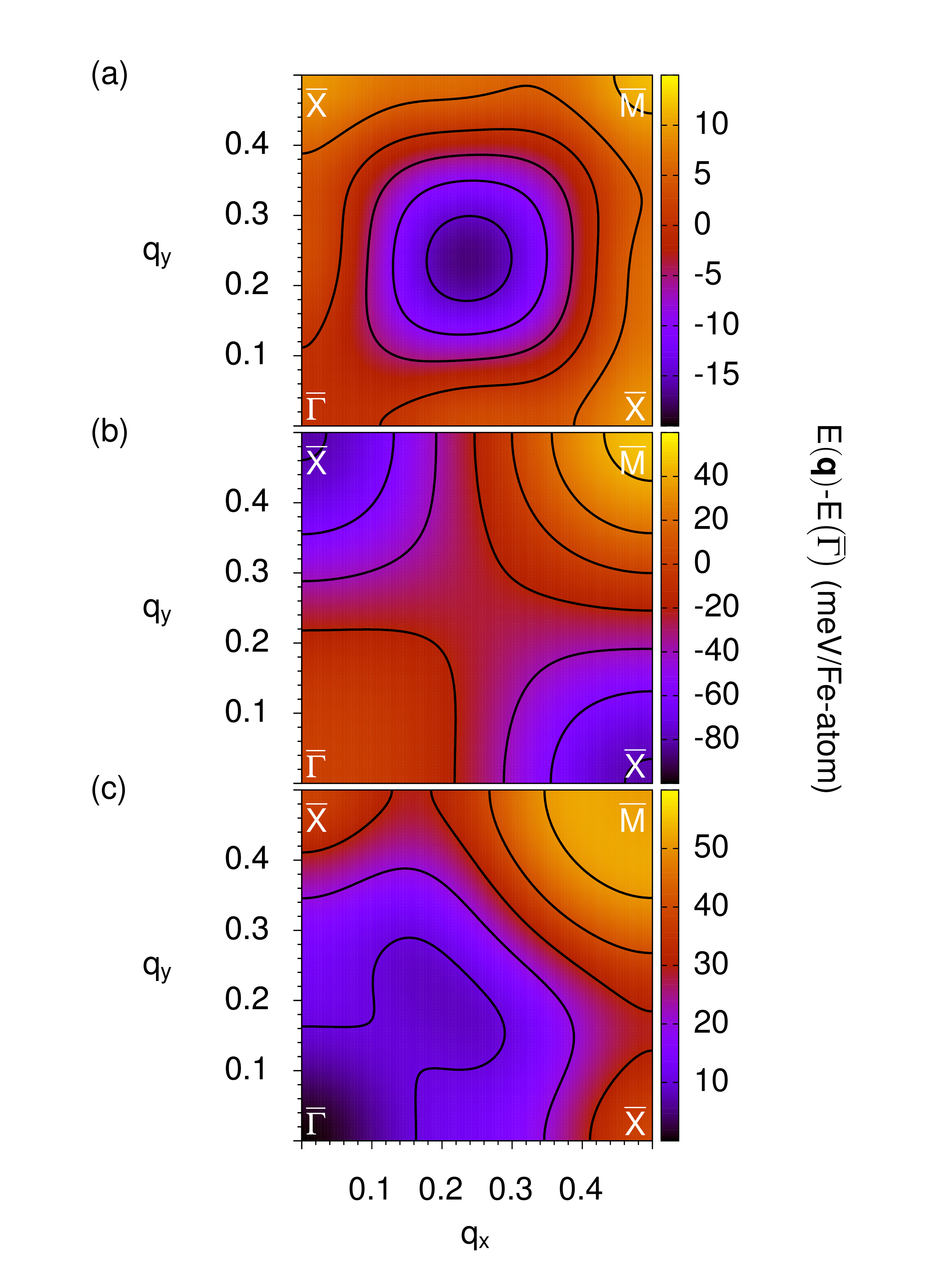}
  \caption{Energy dispersion of flat spin spirals $ E (\textbf{q}) $ in one quarter of the 2D BZ obtained based on the atomistic model with the
  exchange constants determined via DFT (cf.~Table \ref{tab:J}) for (a) Mo/Fe/W(001) (a), Ru/Fe/W(001) (b) and Pd/Fe/W(001). High symmetry points of the
  2D BZ are indicated.}
	\label{fig:extrapolierte_energielandschaft}
\end{figure}

  As discussed in the previous section it is in particular the exchange interaction in the Fe layer which can be strongly tuned by the $4d$ overlayers. 
  Therefore, it is interesting to compare the exchange interaction in Mo/Fe/W(001), Ru/Fe/W(001) and Pd/Fe/W(001) by calculating 
  the energy landscape $ E(\textbf{q}) $ for flat spin spirals in the Brillouin zone 
	[Fig.~\ref{fig:extrapolierte_energielandschaft} based on the determined exchange constants of Table \ref{tab:J}]. 
 
	 From Fig.~\ref{fig:extrapolierte_energielandschaft}~(a) the strong anisotropy in the exchange energy is 
	clearly recognizable for Mo/Fe/W(001). 	 
	 Also in the two-dimensional BZ, the energy minimum lies on the high-symmetry line between $ \overline {\Gamma} $ 
	and $ \overline {\rm M} $ point consistent with the spin spiral energy dispersions shown in Fig.~\ref{fig:film_ss}(a).
   In contrast $E({\mathbf q})$ is isotropic in the vicinity of the $ \overline {\Gamma} $ point for Ru/Fe/W(001) and 
	Pd/Fe/W(001) as seen from the contour lines in Fig.~\ref{fig:extrapolierte_energielandschaft}~(b,c).
	
  The directional anisotropy increases with the length of the spin spiral vector \textbf{q}.
	Close to the \textbf{q} point of the energy
  minimum in Mo/Fe/W(001), i.e.~the $ 90^{\circ}$ spin spiral along $\overline{\Gamma {\rm M}}$, there is a 
	saddle point in Ru/Fe/W(001) and a plateau in Pd/Fe/W(001). 
  The different behavior at this point also highlights the transition of the exchange interaction between the three systems.
  
	These results give an insight, how the exchange interaction can be changed by the additional layer. The energy landscape behaves similar to a tarpaulin, which is fixed at four points. If the heights of the four points are varied against each other, the shape of the landscape changes. In our case this means, we control the shape of the exchange energy by varying the energy differences between the three high symmetry points, 
	as well as the $ 90^{\circ}$ spin-spiral state in $\overline{\Gamma {\rm M}} $-direction.

 	\section{Conclusions}\label{Kap: Zusammenfassung}
 	
  We have demonstrated that it is possible to drastically change the magnetic interactions and the magnetic order 
	in Fe/W(001) by hybridization with a 
  4\textit{d} transition-metal overlayer. Since the nearest-neighbor exchange interaction is relatively small in all studied systems a frustration of 
  exchange and DM interaction occurs. As a result one must go beyond the nearest-neighbor interactions within the 
	atomistic spin model to describe
  the energy landscape accurately. We have studied three systems in detail: Mo/Fe/W(001), Ru/Fe/W(001), 
	and Pd/Fe/W(001) which are all interesting candidates for future experimental studies. 
	
	In Mo/Fe/W(001) the nearest-neighbor exchange and the Fe magnetic moment
	is very small and a spin spiral energy minimum is obtained
  due to a dominating third nearest-neighbor exchange constant. In Ru/Fe/W(001) the antiferromagnetic 
	next-nearest exchange  
  dominates and a row-wise antiferromagnetic state is predicted to be the ground state,
  however, a spin spiral driven by DM interaction is very close in energy. 
	
	Surprisingly, Pd/Fe/W(001) is similar to the Fe monolayer on Ir(111) in
  terms of the exchange interactions and strength of DM interaction 
	only the magnetocrystalline anisotropy energy is significantly larger. 
	Therefore, it would be extremely interesting to find out experimentally whether it shows a conventional
	ferromagnetic ground state or  a similarly 
	intriguing spin structure as the nanoskyrmion lattice reported for Fe/Ir(111)~\cite{Heinze2011}. 
	While we have
	not performed explicit calculations for Rh/Fe/W(001) our results from the Rh/Fe/2W quadlayer
	strongly suggests a spin spiral ground state with a period on the order of 1.5~nm in this system.
   	
 	\begin{acknowledgments}
	  We thank Gustav Bihlmayer for valuable discussions.
  	We gratefully acknowledge computing time at the supercomputer 
 		of the North-German Supercomputing Alliance (HLRN).
 		This project has received funding from the European Unions
 		Horizon 2020 research and innovation programme under grant agreement
 		No 665095 (FET-Opten project MAGicSky).
 	\end{acknowledgments}
 	
 	\bibliography{references}
	
	\end{document}